# A Few Binary Star Puzzles for Roberto on the Occasion of His Birthday

George W. Preston

Carnegie Observatories
813 Santa Barbara Street
Pasadena, California, USA
gwp@ociw.edu

**Abstract:** Radial velocity observations accumulated during the past 16 years are used to derive a preliminary orbit for the CEMP star CS 22881-036. The velocity amplitude is very small. No velocity variation is found for three additional CEMP stars observed over roughly the same time interval. Searches for companions of two CEMP double-lined spectroscopic binaries and of the RR Lyrae star TY Gruis are reviewed. A disparity between the period distribution of disk carbon-star binaries and that of their parent population of normal binaries can be attributed qualitatively to a decline in accreted mass with increasing binary separation. Finally, possible reasons for failure to find expected companions of CEMP stars are discussed.

## 1 Radial Velocity Observations of Three Main Sequence CEMP Stars

Making repeated radial velocity observations of stars that don't change much is a boring activity, as I can affirm from long experience. Sixteen years ago I began to monitor the radial velocities of three CEMP main sequence stars during a search for binaries among the blue metal-poor (BMP) stars (Preston et al. 1994; Preston & Sneden 2000), expecting to make a modest addition to McClure's list of carbon-star binaries (McClure & Woodsworth 1990; McClure 1997). When Preston & Sneden (2001) reported the negative results for these CEMPs, no one paid much attention, understandably perhaps, because of the small numbers of observations (about 12 to 14 per star), the small numbers of epochs (7 in all three cases), and the relatively short duration of the survey (~2500 days) compared to MClure's longer orbital periods. Undeterred, I have continued to make observations through August 2008. These observations are listed in Table 1 and are displayed as plots of radial velocity (RV) *versus* Julian Date (JD) in Figure 1. The radial velocities were derived by cross-correlation of flattened, normalized du Pont echelle spectra ($R \sim 25\,000$) of the program stars with a high S/N template constructed from spectra of one of them, CS 22898-027 for which a radial velocity was calculated from some 200 unblended lines measured with the IRAF package *splot*. The spectra contain 13 overlapping orders that cover the spectral region $\lambda\lambda$ 4000 – 4600 Å. Radial velocities of BMP standard stars measured concurrently were cross-correlated with the template used for the BMP survey of Preston & Sneden (2000).

After noticing a possible small decline in the RV of CS 22881-036 in 2001, I altered my observing schedule to include two observing runs per year in 2003, 2004, 2006, and 2008, and found evidence for a periodicity slightly in excess of one year. My (usual) annual observing trips to Chile act to produce alias periods as noted in Preston & Sneden (2000). This sampling bias is responsible for the apparent slow decline of RV for CS 22881-036 in Figure 1. The relatively large velocity changes between pairs of observing runs in 2003 (JD ~ 2452850) and 2008 (JD ~ 2454650) eliminated the long apparent period. The observations and a radial-velocity curve based on the provisional elements in Table 2 are shown in Figure 2. CS 22881-036 appears to be the primary of a spectroscopic binary with period of 377.5 days and modest eccentricity, $e$ = 0.2. This result conforms to the expectations of the McClure paradigm, stated briefly here for use elsewhere in this paper: all carbon & s-process enriched MS/SG/RGB stars are created by mass transfer from AGB binary companions. Examination of Figure 2 prompts a word of caution. The amplitude of the observed velocity variation is very small. This is illustrated in Figure 3, a histogram of RV semi-amplitudes $K_1$ for CS 22881-036 and for the McClure sample. The $K_1$ value for CS 22881-036 is 50% smaller than the smallest $K_1$ value in the entire McClure sample, i.e., the apparent orbit of CS 22881-036 is not typical of the original sample of stars that define the McClure paradigm.

What can be said about the other two CS stars in Figure 1 is summarized in Table 3, where I have listed the standard deviations of radial velocities separately for observations made with two different detectors used with the du Pont echelle during the course of the survey: the 2D-FRUTTI detector (Shectman 1984) from 1990 through 1999, and a CCD (www.lco.cl/lco/telescopes-information/irenee-du-pont) from 2001 through 2008. There is no significant difference in the precision achieved by use of the two detectors. The standard deviations (SD) are calculated from the $n$ observations for each of the two epochs listed in the third and fifth columns Table 3. During this survey six BMP stars with apparently constant RVs listed in Table 4 of Preston & Sneden (2000) were observed 87 times. The average SD for these, 0.46 kms$^{-1}$, serves as an approximation for the errors of the CEMP RVs. Two features of Table 3 deserve comment. First, the large (1.09 kms$^{-1}$) standard deviation of RVs for CS 22881-036 obtained after 2000 is certainly not an estimator of precision; it arises as a consequence of the decline in RVs measured after 2000 that is evident in Figure 1. Second, the SDs for CS 22880-074 and CS 22898-027 are indistinguishable from those of the radial velocity standards, i.e., there is no evidence of variability in the RVs of these two stars over a time interval of 16 years. If yet-undetected binary variations are present in the data, their $K_1$ values are likely to be smaller than 0.5 kms$^{-1}$, i.e., more than 4 times smaller than that of CS 22881-036, so these stars would lie even farther outside of the bounds of all previous experience concerning carbon-star binaries.

Finally, to complete the exhibit of CEMP main sequence stars that appear to display constant radial velocity, I include in Figure 1 a plot of radial velocity data for LP 706-7 assembled from the literature (Norris et al. 1997; Lucatello et al. 2005; Aoki et al. 2008). Though not as richly sampled as the other three stars, LP 706-7 is a likely candidate for constant velocity in need of more observations.

## 2 What we might learn from study of the DLSBs among the CEMP stars

CS 22964-161 is the primary of a metal-poor ([Fe/H] = -2.4) double-lined spectroscopic binary (hereafter DLSB) (Thompson et al. 2008). This star issues a clear warning that binarity, by itself, does not confirm the presence of AGB-relic companions of CEMP stars. It also announces the potential opportunities afforded by study of CEMP stars in hierarchical triple systems (Evans 1968). By referring to a triple system here I am tacitly invoking the McClure paradigm for CS 22964-161. Note that North et al. (2000) have reported preliminary evidence of a second companion of the Ba dwarf binary HD 48565. However, in that case we cannot know which of the two unseen companions is the white dwarf.

Because the amount of mass accreted from the wind of the putative (AGB) companion of the CS 22964-161 binary is one of the more interesting aspects of this system, I mention some issues that arise in calculation of this quantity. In order to use theoretical AGB winds and calculations of accretion therefrom, we must first establish the dimensions of the system. We do this by appeal to investigations of the stability of triple systems, an aspect of dynamical astronomy pioneered, incidentally, by Torino's own Giuseppi Lodovico Lagrangia (Lagrange 1772). All known triple systems are hierarchical, characterized by $Q = a_3/a_{12}$, the ratio of the orbital semi-major axis of a distant outer member, usually labelled $m_3$, to that of the inner binary ($m_1$, $m_2$).  I use $Q$ to designate this quantity instead of the customary $q$ employed in dynamical astronomy in order to reserve $q$ for its other customary use as mass ratio $q = m_2/ m_1$ in the binary star literature.  Early use of empirical and semi-analytic approximations for the minimum $Q$-value for stability (Heintz 1978; Szebeheley & Zare 1977)) have been supplanted in recent times by minimum $Q$-values for families of binaries generated by numerical simulations, e.g., the extensive tabulations of Kiseleva et al. (1994).  On the observational side a promising empirical method was introduced by Tokovinin (1997).  From the last of these we adopted $Q = 4.6$ ($P_3/P_{12} = 10$ for circular orbits) in our discussion of CS 22964-161.  The literature on which to base a characteristic minimum values of $Q$ is voluminous, as readers may verify by searches with *ADS* or *Google*.

Simplistic treatment of AGB wind accretion produces opposite effects on the inner and outer orbits.  The inner orbit of CS 22964-161 shrank from its initial state to its present state according to Huang (1956), who used McCrea's (1953) "retarding force" that acts on the components of a binary moving in a stationary IS cloud.  Simultaneously, the outer orbit expanded due to mass lost from the system by the AGB wind.  Thompson et al. (2008) calculated the expansion of the orbit and lengthening of the orbital period of the putative AGB relic in the CS 22964-161 system by using the formalism for non-conservative mass transfer developed in section 4.4.2 of Hilditch (2001).  These two processes, accretion by the inner binary and mass loss from the AGB star, both act to increase the $Q$ of the system.

One might suppose that these two changes would render the system more stable, but, this is not necessarily so according to Donnison & Mikulskis (1992), who used numerical simulations to investigate stability of hierarchical triple systems for three mass regimes when $m_3$ is the mass of the outer (AGB) star: $m_3 > (m_1, m_2)$, $m_1 > m_3 > m_2$, and $m_3 < (m_1, m_2)$.  The putative CS 22964-161 system initially lay in the first of these regimes and, after AGB evolution, ended in the third, about which Donnison & Mikulskis noted: "when $m_3$ was the least massive body in the system, the system invariably became unstable through the tendency of $m_3$ to escape from the system altogether."  Evidently, the putative CS 22964-161 triple system remained stable during the ~4.8 Gy main-sequence lifetime of its 1.3 $M_\odot$ AGB antecedent.  It remains for observers to test the prediction of Donnison & Mikulskis by establishing whether or not it was ejected during the ~8 Gy that have elapsed since the AGB evolution of $m_3$.  Ignoring Huang's contraction of the inner binary, the lower limit on the orbital period of the AGB relic is $P > 3700$ d.  If we had used the more restrictive Heintz (1978) stability criterion ($Q > 8$), we would have obtained $P > 8400$ d.  We have continued to make RV observations through August 2008, and find tentative evidence in the most recent observations for a small drift of the center-of-mass velocity.  Measurements must be continued for at least a few more years to make a more definite pronouncement.  Wherever future observations may lead, the system of CS 22964-161 demonstrates that AGB evolution and the dynamics of hierarchical triple systems are inextricably linked.

The foregoing discussion ignores angular momentum carried by the wind of an orbiting AGB star, accretion of which will affect the dynamical evolution of the system.  Unfortunately, calculation of the rate of accretion of angular momentum from such a wind is rendered uncertain according to a conundrum posed by Davies & Pringle (1980).  Here is the conundrum about accretion from an inhomogeneous medium (one in which there are density and/or velocity gradients perpendicular to the flow): in the Bondi-Hoyle model (Bondi & Hoyle 1944), for matter to be accreted at all, cancellation of momentum transverse to the accretion axis is required, and thus *no* angular momentum can be accreted.  The conundrum was revisited by Livio et al. (1986), who concluded that the rate of accretion of angular momentum from an inhomogeneous medium must be significantly lower than previously assumed.  They urged the collection of more observations in order to better understand the accretion process.  Bate & Bonnell (1997) found that changes in proto-binary separations due to accretion from an infalling IS cloud are very sensitive to the assumed specific angular momentum carried by accreted mass.

Recently Soker (2004) discussed the exact situation encountered in the case of CS 22964-161, that is, accretion by a binary moving in the wind of an orbiting AGB star.  He finds that the rate of accretion of angular momentum is sensitive to the mutual inclination of the two orbital planes (there is no *a priori* reason to suppose that they are co-planar). However, the analysis is difficult, Soker was obliged to make a number of plausible assumptions, and there are no error bars on the conclusions. My gloss on this topic is that it is probably not yet possible to calculate the dynamical evolution of triple systems with confidence. Being an aged observer, I feel diffident in making this remark, depending as it does on evaluation of arguments that I scarcely understand. I take refuge in the brusque humor of my countryman, Will Rogers (www.brainyquote.com) who said: "All I know is what I read in the papers …".

One of a kind does not make a class, and unique objects are sometimes put aside and forgotten. However, CS 22964-161 is not alone. Thompson et al. (2008) had barely submitted their manuscript to the Astrophysical Journal when they learned (Jennifer Johnson, private communication) that a second DLSB, CS 22949-008 ([Fe/H] = -2.72, Beers et al. 1992) had been discovered among the metal-poor CEMP stars (Masseron et al. 2008, in preparation). The orbit of this binary is a work in progress, but a preliminary period $P = 335$ d and mass ratio $q = 0.74$ provides a reasonable fit to observations made at five different epochs. Add to these HD 48565 and we do, indeed, have a new class: main-sequence s-process stars in triple systems.

Returning to the warning above, the discovery of two DLSB systems in addition to HD 48565 in the small extant sample of main sequence CEMPs that have been examined for orbital motion demonstrates that detection of a binary companion no longer provides confirmation of the McClure paradigm for CEMP stars. If the masses of the secondaries in the two DLSBs were slightly smaller, we would not have detected them, and application of the McClure paradigm would have led to the erroneous conclusion that they are white dwarfs. What "slightly smaller" means can quantified by reference to the histogram, shown in Figure 4, of the mass ratios of Duquennoy & Mayor (1991) for the F- and G- type stars near the sun. The distribution contains no systems with $q < 0.5$. There is a precipitous drop below $q = 0.65$, which therefore serves as an empirical limit for detection of secondaries, at least among solar-type binaries. This limit is modestly smaller than the mass ratios 0.86 and 0.74 of CS 22964-161 and 22949-008, respectively. For halo turnoff stars $q = 0.65$ corresponds to a lower mass limit for detectable main sequence secondaries $m_2 > 0.52\ M_\odot$.

The foregoing suggests that it would be useful, even prudent, during RV surveys to measure spectrophotometric diagnostics that are sensitive to the presence of a white dwarf, such as those used by E. Bohm-Vitense (1980) in her classic study of the Ba star ξ Cap. In the absence of such tests, how can we rest assured that any particular single-line CEMP binary has a white dwarf companion? The answer to that question is simple: we cannot.

Continuing in this vein it is reasonable to ask whether HE 0024-2523 (Lucatello et al. 2008) and SDSS 1707+58 (Aoki et al. 2008), binaries with orbital periods that are two to three orders of magnitude shorter than those of the bulk of known CEMP stars, experienced the runaway Roche lobe overflow required to produce close white dwarf companions. Perhaps they are merely the ordinary inner binaries of yet-undetected hierarchical triples. This is hardly a preposterous notion. After all, Tokovinin et al. (2006) found that ~1/3 of the 161 normal spectroscopic binaries studied by them reside in hierarchical triples. What if it should turn out that accretion by binaries in an AGB wind is particularly efficient? Were that so, triples among CEMP stars might be the norm. Pushing speculation still further, I only mention that I can conveniently explain the small $K_1$ value of CS 22881-036, discussed in Section 1, by supposing that its companion is a cool main sequence dwarf ($M = 0.12\ M_\odot$) in a modestly inclined orbit ($i \sim 30°$). Were I were ten years younger I would undertake a search for third stars in all close CEMP binaries.

## 3. Where are the long-period carbon star binaries?

Numerous investigations during the past two decades, of which I only mention two here (Han et al. 1995; Theuns et al. 1996), concur that accretion from an AGB wind can produce classical Ba and CH star companions in binaries with orbital periods up to 100 000 days or more, although efficient accretion is generally limited to periods below 10 000 days. On the observational side Boffin & Zacs (1994) found an imperfect but recognizable decline in s-process overabundances with increasing orbital period. In their sample of 17 stars, the two with the longest orbital periods, HD 139195 ($P$ = 5324 d) and HD 202109 ($P$ = 6489 d), both due to Griffin (1991), possess the lowest s-process overabundances. Their results fulfill our demand that AGB pollution must fall below detectable levels in sufficiently wide binaries. An empirical comparison allows us to quantify "sufficiently wide". Compare the distribution of periods of the Ba stars in Figure 2 of Jorissen & Van Eck (2000) with the distribution of periods of F- and G- type main sequence stars in Figure 7 of Duquennoy & Mayor (1991). The former peaks at about 3000 days, and contains no stars with periods as large as 5000 days. Non-conservative mass-loss by AGB winds can only increase orbital periods of detached binaries, so the initial periods of the binaries plotted by Jorissen & Van Eck must have been even shorter than the present, measured values plotted in their diagram – shorter perhaps by a factor of ~2, if the case of CS 22964-161 (Thompson et al. 2008) is a reliable guide. In contrast, the period distribution of the solar-type binaries of Duquennoy & Mayor peaks near 100 000 days. To be sure, Duquennoy & Mayor created this peak of their distribution by adding visual and common proper motion binaries to their spectroscopic sample. For orbital periods up to ~100 000 days they proceeded in this manner as a matter of convenience, not because of failure of the Doppler method. If the Ba and CH binaries are drawn from their parent population of solar-type main sequence binaries, then the bulk of AGB-polluted companions should occur in binaries with orbital periods greater than 10 000 days, but this is not the case. No such binaries have been found. This failure is not a radial-velocity precision problem. The $K_1$ value of a 0.8 $M_{\odot}$ star in a 10 000 day orbit with a 0.5 $M_{\odot}$ white dwarf companion is ~4 kms$^{-1}$, easily detectable by all modern high-resolution spectrographs.

There is a simple explanation for all of this, nevermind that I am unable to quantify the argument. Stellar evolutionary timescales dictate that the overwhelming majority of AGB mass transfers occur during the main sequence lifetimes of their companions. Such main sequence carbon stars near the turnoff can be identified in very wide (long-period) binaries, because small amounts of accreted AGB material are confined to thin surface mass layers by complex interactions of convection, diffusion, and thermohaline mixing (Vauclair 2004; Thompson et al. 2008). Their peculiar spectral signals lie above current survey detection thresholds of appropriate chemical identifiers, the G-band index of Beers et al. (1992), for example. However, stars identified as chemically peculiar in this manner, may not be identified as members of binaries, because their long orbital periods tax the patience of observational astronomers (see, for example, three of the four stars in Figure 1). These appear to fall outside the bounds of the McClure paradigm. The AGB binary enrichment scheme produces a period-dependent distribution of abundance enhancements, only a portion of which, the portion associated with $P < 5\ 000$ d, falls above current detection thresholds. It also skews the orbital period distribution *of identified giant Ba and CH stars*, producing the discrepancy between the period distributions of Jorissen & Van Eck and Duquennoy & Mayor noted above.

The spectral peculiarity indices of such stars gradually diminish during post main-sequence evolution, as their thin surface layers are mixed into deepening convective envelopes on the RGB. Thus, main sequence carbon stars that are polluted by sufficiently distant AGB companions literally disappear during ascent of the RGB. This phenomenon greatly complicates the mapping of main sequence carbon stars onto the RGB by the isochrone method, because the observable mapping parameter, linear density along an isochrone, does not scale with lifetime for *identifiable* extrinsic carbon stars.

The conclusions of the preceding paragraph are based largely on studies of stars in the Galactic disk. However, there are no startling differences between the binary populations of the disk (Duquennoy & Mayor 1991; Udry et al. 1998) and halo (Latham et al. 1988; Latham et al. 1998) with respect to various group properties: binary fraction, proportion of double-lined systems, period distribution, and eccentricity distribution. Therefore, I see no reason not to apply generalizations about the binary population of the Galactic disk to that of the Galactic halo.

**4 TY Gruis: a new CEMP species.**

It was bound to happen sooner or later – the discovery of a carbon and s-process rich RR Lyrae star. TY Gruis (CS 22881-071) is such an object. The status of our search for a companion through 2005 is summarized in Figure 13 of Preston et al. (2007). No binary motion was detected during an interval of 400 days. The sensitivity of radial velocity observations to binary motion in this case is compromised not only by the large velocity range of pulsation ($\sim 60$ kms$^{-1}$), but also by substantial modulation of this range during a Blazhko cycle of 68 days.

I have continued observation of TY Gru for three more years and update my results in Figure 5. The RV extrema of the pulsation cycles, defined by only a small fraction of the observations, all lie near the horizontal straight lines drawn at +20 and -46 kms$^{-1}$: there is no convincing evidence for acceleration of TY Gruis in the RV data collected to date. I have not yet constructed smoothed pulsation velocity curves that can be integrated to derive annual center-of-mass velocities for the seven well-observed epochs shown in Figure 5. I hope, but have not yet demonstrated, that velocities calculated in this manner will more accurately track the center-of-mass velocity of this star. Detection of the binary motion of TY Gruis may prove to be difficult.

**5 Conclusions**

In preceding Sections I have described several observational difficulties encountered in the identification of white dwarf companions of CEMP stars. Binary orbital motion alone no longer suffices as proof, and as yet there is no verification of third bodies in either of the two known DLSB systems, or in short-period systems, i.e., those that experience Roche-lobe overflow (except for the yet-to-be-documented case of HD 48565 mentioned in Section 2 above).

There are a number of more or less plausible reasons for failure to find RV variations in the stars of Section 1 or in the DLSB's of Section 2. I list those of which I am aware below.

(1) The periods of SLSBs (e.g., those in Section 1) could be very long and the stars have not been observed long enough or with requisite precision. Detection of very long periods can be remedied simply by patience. The observed meter-per-second variations of many stars induced by planets (Quirrenbach 2006) assure that precision is not a problem for orbital periods up to $P \sim 100\,000$ d.

(2) Binaries with orbital velocity ranges comparable to intrinsic spectral line-width and with mass ratios (hence luminosity ratios) near unity will remain undetected unless special methods are employed: these are the "line-width binaries" of Griffin (1985). Duquennoy & Mayor (1991) identified six such systems in their survey of F- and G- type stars.

(3) Small orbital inclination can always be invoked to explain individual cases, like those discussed in Section 1 above. A statistical test of this hypothesis could be made persuasive by enlarging the sample and, particularly, by improving the precision in RVs, thereby reducing limits on $K_1$ values and the limits on orbital inclinations calculated from them.

(4) Ejection of white dwarfs from initial triple systems following AGB evolution, as predicted by Donnison & Mikulskis (1992), offers a possibility of a different sort. At present this explanation is of the nature of an hypothesis to be tested. It constitutes a most interesting intersection of AGB physics and dynamical astronomy that will attract increasing attention as the number of test cases accumulates. Because halo triples have ages comparable to the age of the Galaxy, they provide two constraints on stability. The main sequence lifetime of the AGB companion establishes how long the initial triple was stable. The time elapsed from AGB evolution to the present attests to the stability of any triple that has survived to the present epoch.

(5) In more speculative vein, imaginative folks may consider the possibility of an unanticipated disruptive process that leads to ejection of an unconventional source of s-process enrichment: for example, the scheme involving an accretion-induced collapse supernova presented in section 10.3 of Cohen et al. (2003).

(6) Finally, for completeness sake, there lurks the possibility that some CEMP stars "were just made that way". This is one convenient way to explain the low binary frequency (2 out of 32) among the Ba and gCH stars in the globular cluster ω Centauri reported by Mayor et al. (1996). And my friend and colleague Andy McWilliam, noting the large s-process abundances encountered in the Sagittarius dwarf spheroidal galaxy (McWilliam & Smecker-Hane 2005), is willing to contemplate the possibility that one or another of my stars, born in such a place where generations of AGB ejecta polluted an ISM between starbursts, subsequently wandered into the Milky Way halo. I see no harm in making room here for this bit of heresy.

**Figure Captions**

Figure 1  Radial velocity versus Julian date for four CEMP main sequence stars: CS 22880-074 (upper left), CS 22881-036 (upper right), CS 22898-027 (lower left), and LP 706-7 (lower right).  Filled and open circles for the three CS stars denote observations made with 2D-FRUTTI and CCD detectors, respectively, as discussed in the text.   For LP 706-7, filled circles, open circles, and the filled box denote observations by Norris et al. (1997), Lucatello et al. (2005), and Aoki et al. (2008), respectively.

Figure 2  Radial velocity observations and a provisional radial velocity curve for CS 22881-036 calculated by use of the orbital elements in Table 2.  Symbols have the same meanings as those for CS 22881-036 in Figure 1.

Figure 3  Histogram of $K_1$ values for the McClure gCH star and Ba stars(white bars) and CS 22881-036 (black bar).

Figure 4  Distribution of mass ratios in the DLSB sample of Duquennoy & Mayor (1991)

Figure 5  Radial velocities of TY Gruis in the years 2003 -2008

Figure 1 (upper left)

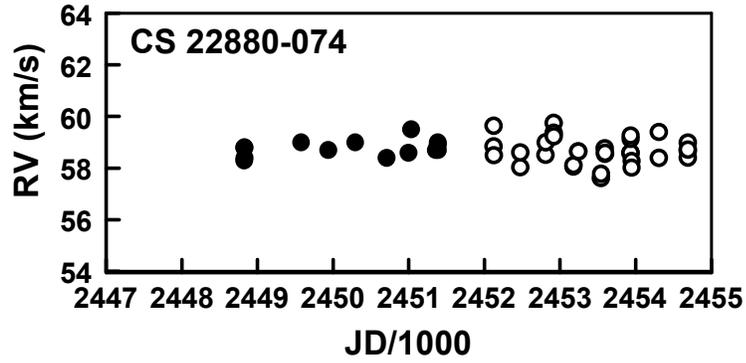

Figure 1 (upper right)

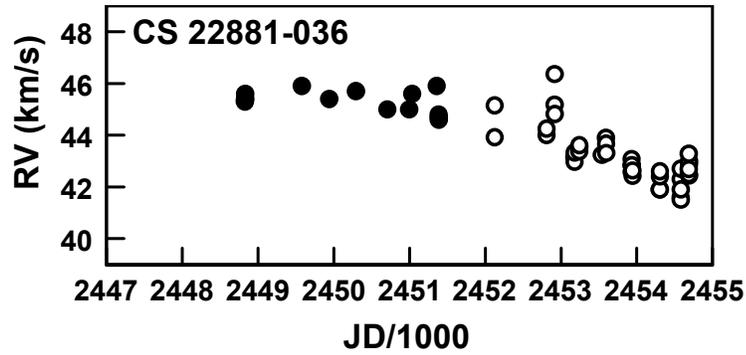

Figure 1 (lower left)

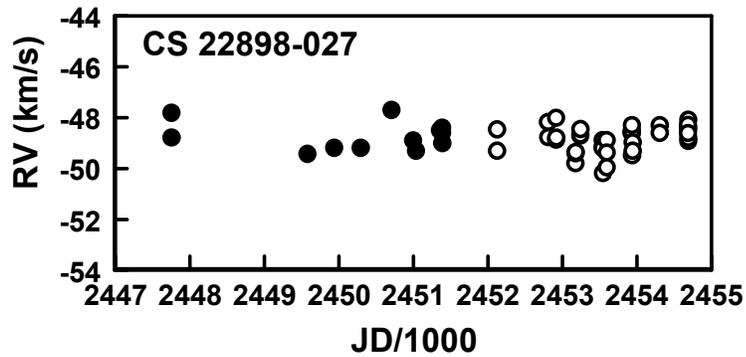

Figure 1 (lower right)

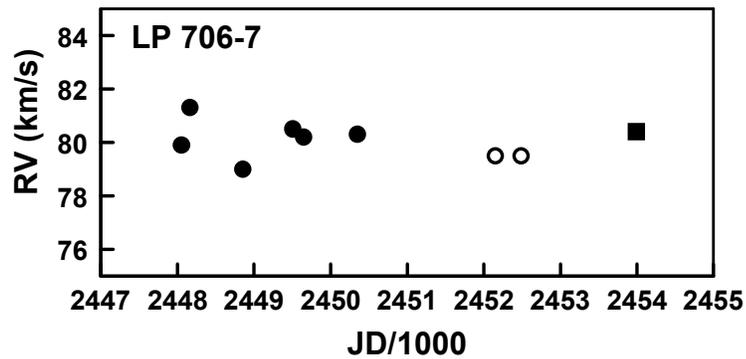

Figure 2

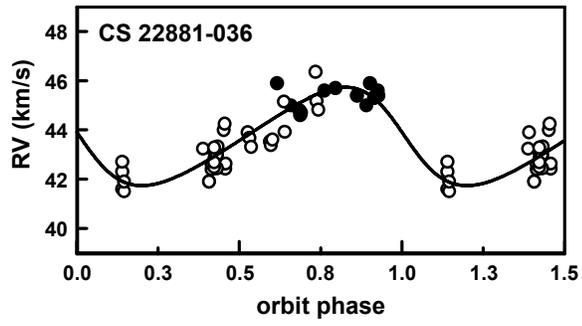

Figure 3

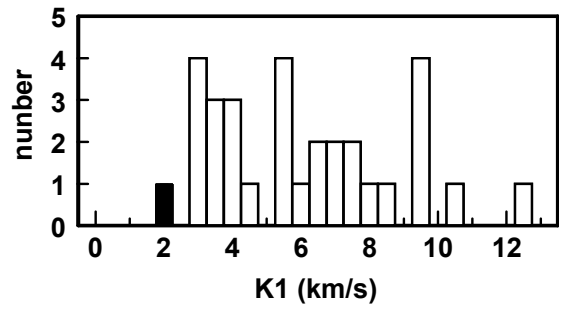

Figure 4

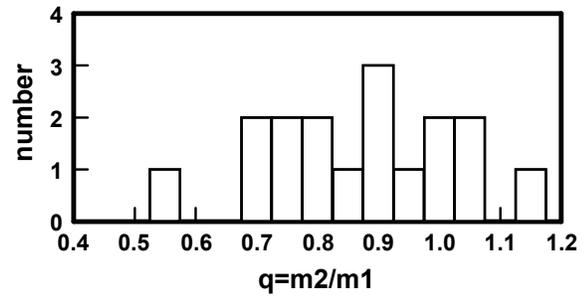

Figure 5

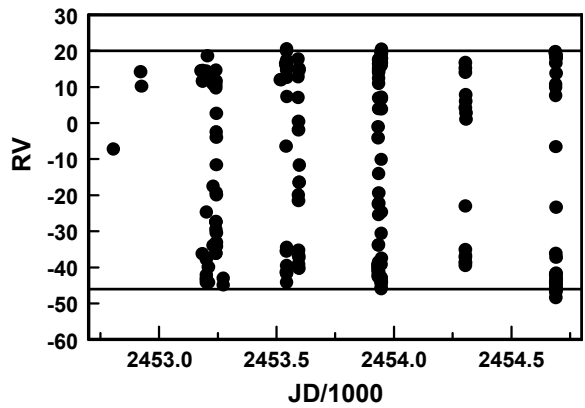

**Table 1  Radial velocity observations of three CEMP stars**

| JD-2400000 | RV(kms$^{-1}$) | JD-2400000 | RV(kms$^{-1}$) | JD-2400000 | RV(kms$^{-1}$) |
|---|---|---|---|---|---|
| CS 22880-074 | | CS 22881-036(cont'd) | | CS 22898-027 | |
| 48825.810 | 58.3 | 48830.730 | 45.4 | 47750.700 | -47.8 |
| 48827.640 | 58.8 | 49576.740 | 45.9 | 47752.600 | -48.8 |
| 48829.590 | 58.8 | 49939.740 | 45.4 | 49576.720 | -49.4 |
| 48830.670 | 58.4 | 50292.680 | 45.7 | 49937.720 | -49.2 |
| 49576.710 | 59.0 | 50706.750 | 45.0 | 50292.640 | -49.2 |
| 49937.710 | 58.7 | 50996.800 | 45.0 | 50706.730 | -47.7 |
| 50292.630 | 59.0 | 51035.740 | 45.6 | 50996.790 | -48.9 |
| 50707.720 | 58.4 | 51358.830 | 45.9 | 51035.730 | -49.3 |
| 50996.780 | 58.6 | 51384.710 | 44.8 | 51358.810 | -48.5 |
| 51033.580 | 59.5 | 51385.690 | 44.6 | 51384.670 | -48.4 |
| 51358.790 | 58.7 | 51386.700 | 44.7 | 51387.670 | -49.0 |
| 51384.660 | 58.9 | 52122.760 | 45.2 | 51386.680 | -48.4 |
| 51385.650 | 59.0 | 52123.730 | 43.9 | 52122.656 | -49.3 |
| 51386.670 | 58.7 | 52808.820 | 44.0 | 52123.764 | -49.3 |
| 52121.660 | 58.8 | 52809.790 | 44.3 | 52808.756 | -48.2 |
| 52122.620 | 59.6 | 52915.634 | 46.4 | 52809.747 | -48.8 |
| 52123.750 | 58.5 | 52916.675 | 45.2 | 52915.537 | -48.9 |
| 52474.740 | 58.0 | 52918.653 | 44.8 | 52916.581 | -48.0 |
| 52474.800 | 58.6 | 53175.809 | 43.2 | 52918.555 | -48.8 |
| 52808.690 | 58.5 | 53178.770 | 43.0 | 53175.770 | -49.8 |
| 52809.720 | 58.6 | 53179.730 | 43.3 | 53176.700 | -49.4 |
| 52915.490 | 59.8 | 53240.692 | 43.5 | 53179.820 | -49.4 |
| 52916.540 | 59.4 | 53241.676 | 43.4 | 53240.604 | -48.7 |
| 52918.503 | 59.2 | 53243.596 | 43.6 | 53241.648 | -48.7 |
| 53175.725 | 58.1 | 53540.840 | 43.2 | 53242.684 | -48.5 |
| 53178.725 | 58.1 | 53592.790 | 43.9 | 53539.740 | -48.9 |
| 53240.556 | 58.7 | 53594.680 | 43.7 | 53540.700 | -49.1 |
| 53241.596 | 58.7 | 53596.700 | 43.3 | 53541.760 | -49.2 |
| 53540.690 | 57.6 | 53931.868 | 43.1 | 53544.670 | -50.2 |
| 53542.660 | 57.8 | 53931.890 | 42.8 | 53592.720 | -48.9 |
| 53592.640 | 58.8 | 53933.845 | 42.8 | 53594.800 | -49.4 |
| 53594.760 | 58.6 | 53933.864 | 42.6 | 53597.560 | -50.0 |
| 53596.610 | 58.6 | 53944.765 | 42.4 | 53931.805 | -48.6 |
| 53931.836 | 58.6 | 53944.786 | 42.6 | 53931.815 | -48.6 |
| 53931.851 | 58.6 | 54303.774 | 41.9 | 53933.713 | -48.5 |
| 53933.685 | 59.1 | 54303.793 | 41.9 | 53933.724 | -48.3 |
| 53933.698 | 59.3 | 54306.832 | 42.4 | 53935.841 | -49.5 |
| 53945.798 | 58.3 | 54306.849 | 42.6 | 53944.614 | -49.0 |
| 53945.812 | 58.0 | 54580.874 | 41.6 | 53944.627 | -49.3 |
| 54308.583 | 59.4 | 54580.889 | 42.3 | 54306.794 | -48.3 |
| 54308.597 | 58.4 | 54580.903 | 42.7 | 54306.802 | -48.6 |
| 54689.724 | 58.7 | 54582.878 | 41.5 | 54691.661 | -48.1 |
| 54689.735 | 58.4 | 54582.896 | 41.9 | 54687.778 | -48.5 |
| 54690.670 | 58.7 | 54688.839 | 42.8 | 54690.702 | -48.9 |
| 54690.682 | 59.0 | 54688.821 | 42.9 | 54691.651 | -48.4 |
| | | 54690.889 | 42.4 | 54690.693 | -48.8 |
| CS 22881-036 | | 54690.870 | 43.0 | 54687.769 | -48.3 |
| 48825.850 | 45.3 | 54687.840 | 42.5 | 54687.759 | -48.6 |
| 48827.740 | 45.4 | 54687.796 | 42.7 | | |
| 48829.830 | 45.6 | 54687.819 | 43.3 | | |

**Table 2  Provisional orbital elements for CS 22881-036**

| Element | Value |
| --- | --- |
| JDo | 2448484.0 |
| $V_0$ | 43.7 |
| $K_1$(kms$^{-1}$) | 2.0 |
| e | 0.2 |
| $\omega$(deg) | 85 |
| P(days) | 378 |

**Table 3  Standard deviations of radial velocities for three CEMP stars observed from 1990 through 2008**

| Star | SD(2DF) kms$^{-1}$ | n | SD(CCD) kms$^{-1}$ | n |
| --- | --- | --- | --- | --- |
| CS 22880-074 | 0.30 | 14 | 0.54 | 27 |
| CS 22881-036 | 0.41 | 14 | 1.09 | 42 |
| CS 22898-027 | 0.53 | 12 | 0.53 | 24 |